\newif\ifmnras
\mnrastrue
\ifmnras
    	\documentclass[apj,numberedappendix]{emulateapj}
\else
	\documentclass[apj,numberedappendix]{emulateapj}
\fi

\usepackage{graphics,epsf}
\usepackage{amsmath}                
\usepackage{amsfonts}               
\usepackage{amssymb}                
\usepackage{epsfig}                 
\usepackage{float}
\usepackage{color}

\def \eV{~\rm{eV}}

\def \cm{~\rm{cm}}
\def \s{~\rm{s}}
\def \km{~\rm{km}}

\def \K{~\rm{K}}
\def \g{~\rm{g}}

\def \erg{~\rm{erg}}

\def \yr{~\rm{yr}}

\ifmnras
	\def \aap{A\&A}
	
	\def \apj{ApJ}
	\def \apjl{ApJ}
	\def \apjs{ApJS}

	\def \mnras{MNRAS}
	\def \na{New Astron.}
    \def \nar{New Astron. Rev}
    
	\def \pasa{Publ. Astron. Soc. Australia}

\fi

\usepackage{xcolor}
\definecolor{redak}{rgb}{0.9,0.15,0.05}

\begin{document}

\ifmnras

\title{Radiating the hydrogen recombination energy during common envelope evolution}
\author{Noam Soker\altaffilmark{1,2}, Aldana Grichener\altaffilmark{1}, \& Efrat Sabach\altaffilmark{1}}

\altaffiltext{1}{Department of Physics, Technion -- Israel Institute of Technology, Haifa 32000, Israel; soker@physics.technion.ac.il; aldanag@campus.technion.ac.il; efrats@physics.technion.ac.il}
\altaffiltext{2}{Guangdong Technion Israel Institute of Technology, Shantou 515069, Guangdong Province, China}

\begin{abstract}
By using the stellar evolution code \texttt{MESA} we show that most of the hydrogen recombination energy that is released as the envelope expands during a regular common envelope evolution (CEE), namely, the initial dynamical phase or plunge-in phase, is radiated, and hence increases substantially the stellar luminosity.
 Only about ten per cent of the hydrogen recombination energy might be used to remove the envelope.
We show that the key property of energy transport is that when convection becomes inefficient in the outer parts of the envelope, where the ionization degree of hydrogen falls below about 30 per cent, photon diffusion becomes very efficient and removes the recombination energy.
The expanding envelope absorbs most of the gravitational energy that is released by the spiraling-in process of the secondary star inside the common envelope, and so it is the hydrogen recombination energy that is behind most of the luminosity increase of the system.
The recombination energy of hydrogen adds only a small fraction of the energy required to remove the common envelope, and hence does not play a significant role in the ejection of the envelope.
\end{abstract}



\section{INTRODUCTION}
\label{sec:intro}

Two open questions concerning the energetics of the common envelope evolution (CEE) that attracted attention in recent years are the question regarding the role of hydrogen and helium recombination energy in facilitating envelope ejection, and the question regarding the role of the gravitational energy from the mass that the companion (secondary) star accretes from the giant envelope.
The quest for extra energy sources comes from the results of hydrodynamical numerical simulations that show that it is not straightforward to remove the common envelope in a short time by using only the orbital energy of the in-spiraling core-companion system (e.g., \citealt{Ohlmannetal2016, Ohlmannetal2016b, Staffetal2016a, Staffetal2016MN8, NandezIvanova2016, Kuruwitaetal2016, IvanovaNandez2016, DeMarcoIzzard2017, Galavizetal2017, Iaconietal2017, Iaconietal2018, MacLeodetal2018}, limiting the list to the last three years).

Over a long time and at the termination of the CEE other sources might play a role, including excitation of p-waves \citep{Soker1993}, interaction of the core-secondary system with a circumbinary disk
(e.g., \citealt{Kashisoker2011MN, Kuruwitaetal2016}), envelope inflation followed by vigorous pulsation \citep{Claytonetal2017}, the stellar luminosity itself that exerts force on dust (e.g., \citealt{Soker2004, GlanzPerets2018}), and jets that are launched by the secondary star (for a full list of processes see \citealt{Soker2017final}).

Jets might also play a role in helping envelope removal at earlier CEE phases (\citealt{Soker2016Rev} for a review and, e.g., \citealt{MorenoMendezetal2017}, \citealt{LopezCamaraetal2018} and \citealt{ShiberSoker2018} for recent hydrodynamical simulations). A key question for jet activity is whether accretion disks or belts are formed
(e.g., \citealt{MacLeodRamirezRuiz2015NS, MacLeodRamirezRuiz2015disk, MurguiaBerthieretal2017}) and whether jets can allow a high mass accretion rate (e.g., \citealt{Shiberetal2016, Chamandyetal2018}). In extreme cases, a neutron stars that launches jets inside a giant envelope might lead to a violent event termed common envelope jets supernova (CEJSN; \citealt{SokerGilkis2018}), or CEJSN impostor \citep{Gilkisetal2018}.

A stronger debate centers around the role of recombination energy, whether it is important for envelope removal
(e.g., \citealt{IvanovaNandez2016, Kruckowetal2016, NandezIvanova2016, Ivanova2018} and references therein), or not (e.g. \citealt{Sabachetal2017, Gricheneretal2018} and references therein).
The observations of Balmer emission lines as well as an effective temperature of $\simeq 6000 \K$ at early times from some intermediate luminosity optical transients (ILOTs; e.g., \citealt{Munarietal2002, Smithetal2016}) show that some recombination energy does leak out from expanding envelopes.
The escape of recombination energy is more pronounced in Type IIp supernovae that show a plateau in their light curve (e.g., \citealt{Galbanyetal2016}). It is the leakage of recombination energy that maintains a more or less constant luminosity during the plateau phase (e.g., \citealt{DessartHillier2011, Faranetal2018}).
As for ILOTs, \cite{MacLeodetal2017M31} discuss a model for the transient event M31LRN 2015, where the outburst is a dynamically driven ejecta at the onset of a CEE phase with a progenitor of mass $3-5.5 M_\odot$. In their model of this CEE event the recombination energy of the ejected gas is only $< 2 \%$ of its kinetic energy. Indeed, recombination energy becomes less efficient as the star becomes more massive.
\cite{Pejchaetal2016} suggest that mass loss through the outer Lagrange point before the secondary star enters the giant envelope can also power the radiation of ILOTs, as well as remove mass before the onset of the CEE.

In a recent paper \cite{Ivanova2018} claimed that hydrogen recombination indeed plays an important role in common envelope ejection and criticized the opposing claim we made in an earlier paper \citep{Gricheneretal2018}.
\cite{Ivanova2018} further argued that the process we study in \cite{Gricheneretal2018} applies to the self-regulated CEE phase though this assertion is not true as we studied the plunge-in (dynamical) phase when the envelope rapidly expands.
In that respect we differ from calculations of the convective energy transport during the long self-regulated CEE phase (e.g., \citealt{MeyerHofmeister1979, Podsiadlowski2001, Ivanovaetal2015}).
 \cite{Ivanova2018} further argued that convection and radiation cannot transport much of the recombination energy out.
In the present study we raise arguments to the contrary.

\section{Envelope inflation}
\label{sec:evolution}

We describe here the relevant properties of the evolution from our earlier paper (\citealt{Gricheneretal2018}, where all details can be found), and present new data from that simulation.

We run the stellar evolution code \texttt{MESA} (Modules for Experiments in Stellar Astrophysics), version 9575 (\citealt{Paxtonetal2011, Paxtonetal2013, Paxtonetal2015,Paxtonetal2018}) to follow the evolution of a star with an initial mass (on the zero age main sequence) of $M_{\rm 1,ZAMS}=2 M_{\odot}$ and with a metallicity of $Z=0.02$. When the star becomes an asymptotic giant branch (AGB) star with a mass of $M_{\rm 1,AGB}=1.75 M_{\odot}$ and a radius of $R(t=0) = 250R_\odot$ we inject energy into the envelope to mimic a companion star of mass $M_2=0.3 M_\odot$ that spirals-in from the surface to an orbital separation of $a=50 R_\odot$ in $1.7 \yr$.
We deposit the energy in the envelope zone that satisfies $50 R_{\odot}<r< 120 R_{\odot}$ with a constant energy per unit mass, and with a total power of $q = 4.5 \times 10^{37} \erg s^{-1}$ (see our earlier paper for details of the entire scheme).
We run \texttt{MESA} in its hydrostatic module. Because the envelope inflation time of $1.7 \yr$ is longer than the dynamical time of the star of $0.3 \yr$, even at its larger size, this treatment is justified.

In Fig. \ref{fig:evolution} we present the evolution of the stellar luminosity $L$ (upper panel), the stellar radius $R$ and the radius at which hydrogen is ionized to a degree of $\chi=50\%$, $R_{\rm ion,50}$, and the mass coordinate, $m_{\rm ion,\chi}$ of the zones where hydrogen is ionized to a degree of $\chi=90\%$, $\chi=50 \%$ and $\chi=10\%$.
\begin{figure}
\begin{center}
\vspace*{-6.2cm}
\hspace*{-4.4cm}
\includegraphics[width=0.95\textwidth]{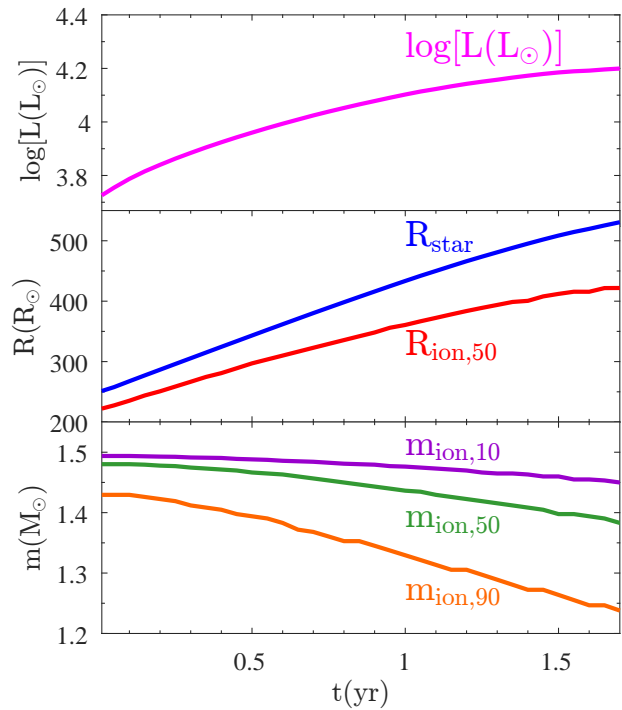}
\vspace*{-6.5cm}
\caption{The evolution of some stellar parameters of the AGB model during the energy injection period.
Upper panel: The stellar luminosity. Middle panel: The radius of the photosphere (blue) and the radius $R_{\rm ion,50}$ where hydrogen is $\chi=50\%$ ionized (red).
Lower panel: mass coordinates of the zones where hydrogen is ionized to a degree of $\chi=10\%$ (purple), $\chi=50 \%$ (green), and $\chi=90\%$ (orange). }
\label{fig:evolution}
\end{center}
 \end{figure}

To better follow the energy transport, in Figs. \ref{fig:structure0}-\ref{fig:structure17} we present relevant quantities as a function of radius in the ionization zone of hydrogen at three times. In the upper panel of each figure we present the luminosity $L(r)$ in logarithmic scale. In the middle panel of each figure we present the adiabatic derivative $\nabla_{\rm ad} = (d\ln T/ d \ln P)_{\rm ad}$ (blue), and the ratio of the local mixing length of the convective cells to the radius, $\Lambda/r$ (red).
In the lower panel we present the ratio $c^2_s/c_p T$, where $c_p(r)$ is the heat capacity at constant pressure per unit mass, $T(r)$ is the temperature, and $c_s(r)$ is the sound speed. We will make use of this ratio in section \ref{subsec:cells}.
\begin{figure}
\begin{center}
\vspace*{-5.50cm}
\hspace*{-2.3cm}
\includegraphics[width=0.7\textwidth]{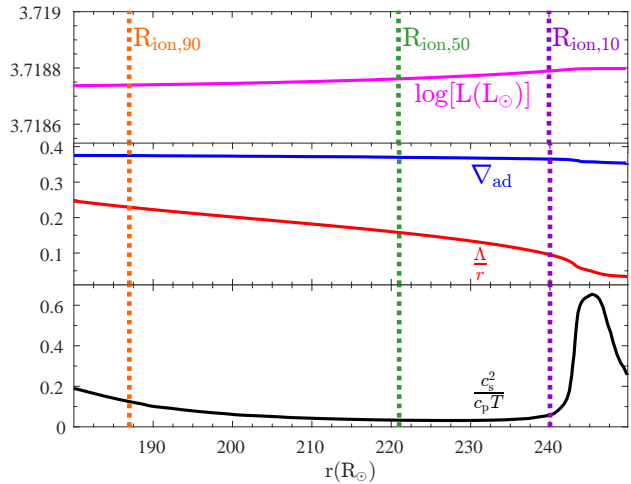}
\vspace*{-5.0cm}
\caption{Profiles of quantities in the outer layers of the envelope of our AGB model at $t=0$ (the beginning of envelope inflation that mimics the plunge-in phase). Upper panel: The luminosity. Middle panel: The adiabatic derivative $\nabla_{\rm ad} = (d\ln T/ d \ln P)_{\rm ad}$ (blue) and the ratio of mixing length to radius $\Lambda/r$ (red). Lower panel: The quantity $c^2_s/c_p T$, where $c_s$ is the sound speed, $c_p$ the heat capacity per unit mass, and $T$ is the temperature.
The orange, green and purple dotted vertical lines mark the zones of $90\%$, $50\%$ and $10\%$ ionization of hydrogen, respectively.
}
\label{fig:structure0}
\end{center}
\end{figure}
\begin{figure}
\vspace*{-4.90cm}
\hspace*{-2.3cm}
\includegraphics[width=0.7\textwidth]{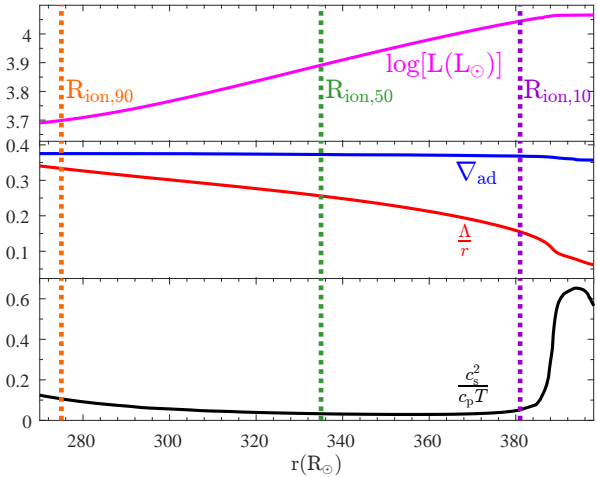}
\vspace*{-5.0cm}
\caption{ Like Fig. \ref{fig:structure0} but at $t=0.8 \yr$.
}
\label{fig:structure08}
\end{figure}
\begin{figure}
\begin{center}
\vspace*{-4.90cm}
\hspace*{-2.3cm}
\includegraphics[width=0.7\textwidth]{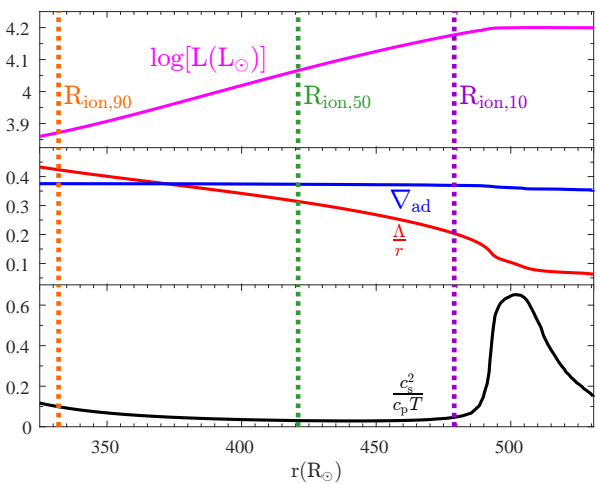}
\vspace*{-5.0cm}
\caption{ Like Fig. \ref{fig:structure0} but at $t=1.7 \yr$.
}
\label{fig:structure17}
\end{center}
\end{figure}

\section{Energy transport}
\label{sec:transport}

We now compare the claims we have previously made \citep{Gricheneretal2018}, that photon diffusion and convection carry out most of the hydrogen recombination energy and that most of the recombination energy is then radiated away, with the counter arguments of \cite{Ivanova2018}.

 We start by a general description of the processes we are about to study. The photons that are emitted by the recombining hydrogen atoms are energetic and deposit energy where they are scattered. The same happens in the Sun, as photons on the solar surface are much less energetic than those in its radiative inner regions. If the envelope does not expand (i.e., it has a very long expansion time), this energy is channeled only into heat, that in turn creates more photons. The number of photons increases as a function of increasing radius. Only if the expansion time of the envelope is shorter, or not much longer, than the time to transport energy out, a large fraction of the thermal energy is channeled to gravitational and kinetic energy. The reason is that in this case the extra pressure that comes from heating by radiation does work on the gas, \textit{before} the energy is transported out. The relation between the energy that is channeled to envelope removal, i.e., to gravitational and kinetic energy, and the energy that is carried out by photon transport depends on the expansion time of the envelope and on the energy transport time. We now present this approximate relation.

We emphasize that we do not claim that recombination energy does not contribute at all to envelope removal, but rather that only a small fraction of the recombination energy is channeled to gravitational and kinetic energy.
Let $t_{\rm tran}$ be the time to transport energy out, both by convection and radiation which operate simultaneously. Namely,
\begin{equation}
t_{\rm tran} < {\rm min} (t_{\rm diffusion}, t_{\rm convection}),
\label{eq:t(tran)}
\end{equation}
where $t_{\rm diffusion}$ is the time to carry energy out from radius $r$ by photon diffusion (radiative diffusion) only, and $t_{\rm convection}$ is the time to carry energy out by convection only.
Let $t_{\rm exp}$ be the expansion time of the envelope. In our case it is about $1.7 \yr$.
The fraction of the photons' energy that can be used to remove the envelope  rather than be radiated away is given approximately by (eq. 2 in \citealt{Sabachetal2017})
\begin{equation}
f_{\gamma} = \frac {t_{\rm tran}}{t_{\rm tran} + t_{\rm exp}} .
\label{eq:Fgamma}
\end{equation}
This relation is based on the derivation of \cite{Arnett1979}, and comes basically from the relation $t^{-1}_{\rm int}=t^{-1}_{\rm exp}+t^{-1}_{\rm tran}$ that comes from their equation (3) \footnote{Where $\tau_{\rm exp}$ of Arnett is $t_{\rm exp}$ here, and $\tau_{\rm dif}$ of Arnett is $t_{\rm tran}$ here.} without nuclear energy production, that in turns comes from the first law of thermodynamics. Here $t_{\rm int}$ is the time scale for the decrease in the internal energy of the expanding envelope as a result of expansion and radiation losses. The energy that is channeled to do work on the expanding envelope is $f_{\gamma}= t^{-1}_{\rm exp}/t^{-1}_{\rm int}$, which is just equation (\ref{eq:Fgamma}).

The implication of equation (\ref{eq:Fgamma}) is as follows. When photons scatter and diffuse out they lose energy. If the envelope does not expand, then the thermal energy is converted to more lower-energy photons that transfer energy out. This is the case in the present day Sun for example. If the envelope expands on a shorter time scale than the energy transport time, i.e., $t_{\rm exp} \ll t_{\rm tran}$, then the thermal pressure does work on the envelope and the photon energy is transfered mainly to envelope removal. In this way the fraction of recombination energy that is radiated away is only $t_{\rm exp} /  t_{\rm tran} \ll 1$ (e.g., \citealt{KasenWoosley2009} for supernovae). In the present study both $t_{\rm tran}$ and $t_{\rm exp}$ are not known to high precision. We will take for $t_{\rm tran}$ either the convection or the radiative diffusion time. Since both operate simultaneously, the transport time in reality is shorter than what we will take in the estimates to come (equation \ref{eq:t(tran)}). For that reason, in this study we actually somewhat overestimate the role of recombination in removing the envelope.

\subsection{The common envelope phase we study}
\label{subsec:phase}

 As seen from Fig. \ref{fig:evolution} the radius substantially increases during the $1.7 \yr$ time period studied here. From the upper panel in both Fig. \ref{fig:structure08} and Fig. \ref{fig:structure17} we see that the luminosity substantially increases with radius in the ionization zone of hydrogen. These fast radius evolution and steep luminosity gradient indicate that we are dealing with the dynamical phase (plunge-in phase, or regular CEE as termed in \citealt{Ivanova2018}). We are not dealing with the self-regulated CEE phase, where the evolution is much slower, e.g., the change of the photosphere radius with time is slow and the variation of luminosity with radius in the envelope is shallow.
We do deal with the dynamical phase, contrary to what \cite{Ivanova2018} attributes to our calculation.

\subsection{Photon diffusion}
\label{subsec:diffusion}

In our earlier paper \citep{Gricheneretal2018} we already noticed that the photon diffusion times from the outer regions of the envelope are short. The expression for the photon diffusion time from a radius $r$ to the photosphere at $R_{\rm ph}$ that we use is $t_{\rm diffusion} (r) \approx 3 (R_{\rm ph}-r) \tau (r) /c$, where $c$ is the speed of light, $\tau(r)=\int_{r}^{R_{\rm ph}} \rho(r^\prime) \kappa(r^\prime) dr^\prime$ is the optical depth at radius $r$ from where the photons start to diffuse, and $\kappa (r^\prime)$ is the opacity. From our graphs there we notice that from the radius $R_{\rm ion, 30}$, namely where hydrogen is ionized to a degree of $\chi = 30 \%$, the photon diffusion time is less than a year. At $t=0$, $0.8 \yr$ and $1.7 \yr$ the photon diffusion times from $R_{\rm ion, 30}$ are $t^{30} _{\rm diffusion}(0)=0.2 \yr$, $t^{30}_{\rm diffusion}(0.8)=0.3 \yr$, and $t^{30}_{\rm diffusion}(1.7)=0.5 \yr$, respectively. All these are substantially shorter than the envelope expansion time of $1.7 \yr$, and we find from equation (\ref{eq:Fgamma}) that $f_\gamma \la 0.2$.
  Even from deeper zones where the hydrogen ionization fraction is $\chi=40 \%$ the diffusion time is shorter than $1.7 \yr$, with values of $t^{40} _{\rm diffusion}(0)=0.6 \yr$, $t^{40}_{\rm diffusion}(0.8)=0.8 \yr$, and $t^{40}_{\rm diffusion}(1.7)=1.7 \yr$.

The inequality $t_{\rm diffusion} (m_{\rm ion,30}) \la 0.5 \yr \ll 1.7 \yr$ during the evolution implies that most of the recombination photons from the outer region $m \ga m_{\rm ion,30}$ can escape just by diffusion, even if convection would have been highly inefficient.
About half of the recombination energy from the shell $m_{\rm ion,40} \la a \la m_{\rm ion,30}$ can also diffuse out as radiation since $t_{\rm diffusion}\simeq t_{\rm exp}$ (by using equation \ref{eq:Fgamma}).
Therefore, photon diffusion from this outer zone, where $t_{\rm tran} < 0.5 \yr$, by itself crudely takes away about a quarter of the recombination energy of hydrogen (by equation \ref{eq:Fgamma}).

This result refutes the conclusion of \cite{Ivanova2018} that `` . . . the amount of recombination energy that can be transferred away by either convection or radiation from the regions where recombination takes place is negligible.''
We reiterate our claim that the usage of the entire recombination energy budget to eject the envelope is unjustified before we even add convective energy transport that removes even more recombination energy from the envelope.
 Using the derived quantities above in equation (\ref{eq:Fgamma}) implies that the fraction of the photons' energy that can be used to remove the envelope is $f_\gamma < 0.2$, with a typical number more likely to be
$f_\gamma \simeq 0.1$.

\subsection{The actual luminosity}
\label{subsec:luminosity}

From Fig. \ref{fig:evolution} we learn that the mass coordinate of $\chi=90\%$ hydrogen ionization, $m_{\rm ion,90}$, decreases during the $1.7 \yr$ by about $0.2M_\odot$.
At the same time $m_{\rm ion,50}$ decreases by about $0.1 M_\odot$, and $m_{\rm ion,10}$ by about $0.05 M_\odot$.
We take the equivalent mass that completely recombines to be $M_{\rm rec}=0.1 M_\odot$, about the decrease in the mass coordinate of $m_{\rm ion,50}$. For a solar composition this amounts to an energy of of $E_{\rm rec,H}=1.8\times 10^{45} \erg$. During the energy injection phase that lasts for $1.7 \yr$ the average luminosity due to recombination alone is $L_{\rm rec,H} = E_{\rm rec,H}/1.7 \yr = 8800 (M_{\rm rec}/0.1M_\odot) L_\odot$.

From the upper panels of Figs. \ref{fig:structure08} and \ref{fig:structure17} we see that the increase in the luminosity within the ionization zone of hydrogen, from about $95 \%$ ionization to the photosphere, is about $\Delta L(0.8)= 6900 L_\odot\simeq0.8L_{\rm rec,H}$ at $t=0.8 \yr$, and $\Delta L(1.7)=8800 L_\odot\simeq L_{\rm rec,H}$ at $t=1.7 \yr$.
Since we do not inject energy in that region, and the expansion of the envelope absorbs energy, the increase in luminosity within that region must be due to recombination energy.

From the equality $0.8 L_{\rm rec,H} \la \Delta L \la L_{\rm rec,H}$ we conclude that the convection according to the mixing length theory, that \texttt{MESA} uses, can arrange itself to transport out most of the extra energy that hydrogen recombination releases. This is in contradiction with the claim of \cite{Ivanova2018} that convection can carry only a negligible fraction of the hydrogen recombination energy.

\subsection{Recombination inside convective cells}
\label{subsec:cells}

In \cite{Sabachetal2017} we followed \cite{QuataertShiode2012} and adopted the following expression for the maximum convective flux
$ L_{\rm max,conv,0}(r) = 4 \pi \rho(r) r^2 c_s^3(r),$
where $\rho (r)$ and $c_s(r)$ are the density and the sound speed at radius $r$,
respectively.
This expression takes the heat content of the gas per unit mass to be $c_s^2$, namely the recombination energy is assumed to be zero.

In our previous paper \citep{Gricheneretal2018} we took a different approach, and examined specifically what happens to the recombination energy when a convective cell moves out.
If the photon diffusion time from radius $r$ to the photosphere is longer than the expansion time (as required if the recombination energy is to be used for envelope removal), so is the photon diffusion time out from convective cells. The reason is that the size of the convective cells is about the size of the mixing length $\Lambda$, which is not much smaller than the radius; we find from Figs. \ref{fig:structure0}-\ref{fig:structure17} that the typical ratio is $\Lambda/r \ga 0.2$ in the relevant zones.
When the convective cell moves outwards, it cools and recombines. If the diffusion time is long, the cell carries all of its recombining energy out.
For a solar composition the specific heat of the gas corresponds to a specific energy of
\begin{eqnarray}
\begin{aligned}
e_{\rm rec}({\rm H}^+) & = 13.6 \frac{X_{\rm H}}{m_H} \eV = 9 \times 10^{12} \erg \g ^{-1}
\\ &
= (30 \km \s^{-1})^2 \equiv \beta c^2_s,
\end{aligned}
\label{eq:beta}
\end{eqnarray}
where $X_{\rm H}$ is the mass fraction of hydrogen and $m_H$ is the hydrogen atomic mass.
For the typical sound speed in the recombination zone, $c_s \simeq 11 \km \s^{-1}$, we find that $\beta ({\rm H}^+) \simeq 7.5$ \citep{Gricheneretal2018}. The maximum convective luminosity when recombination takes place in an expanding envelope, therefore,  can be
\begin{equation}
L_{\rm max,conv,rec}(r) \simeq 4 \pi \beta \rho(r) r^2 c_s^3(r),\;\;{\rm with}\;\beta>>1.
\label{eq:lmaxconv1}
\end{equation}
\cite{Christy1962} gives the maximum convective luminosity as $L_{\rm max,conv,C}(r) \simeq 4 \pi  k_C r^2 e_{\rm rec}({\rm H}^+) c_s(r)$, where he writes that $k_C$ is closer to 0.1 than to 1.With this value of $k_C$, this maximum luminosity becomes $L_{\rm max,conv,C}(r) \simeq 4 \pi  r^2 c^3_s(r)$, as used by \cite{QuataertShiode2012}.

\cite{Ivanova2018} takes the maximum convective luminosity to be $ L_{\rm max,conv,I}(r) = 4 \pi \rho(r) r^2 c_s c_p T \nabla_{\rm ad}$,
where $c_p$ is the specific heat capacity and $\nabla_{\rm ad} =(d \ln T / d \ln P)_{\rm ad} $ is the adiabatic derivative. The ratio of the maximum convective luminosity we use to that of \cite{Ivanova2018} reads
\begin{equation}
\frac {L_{\rm max,conv,rec}}{L_{\rm max,conv,I}} =
\frac{\beta}{\nabla_{\rm ad}} \frac {c_s^2}{c_p T} .
\label{eq:ratioR}
\end{equation}
To estimate the typical value of this ratio we present the values of $ c^2_s/c_p T$ and $\nabla_{\rm ad}$ at three times in Figs.
\ref{fig:structure0}-\ref{fig:structure17}. The small ratio $ c^2_s/c_p T \approx 0.1$ in the relevant region results from the fact that the heat capacity includes the ionization/recombination energy. Inserting these values into equation (\ref{eq:ratioR}) we find that the two maximum convective luminosities are approximately equal
\begin{equation}
\frac {L_{\rm max,conv,rec}}{L_{\rm max,conv,I}} =
\left( \frac{\beta}{20 \nabla_{\rm ad}} \right)
\left( \frac {c_s^2}{0.05 c_p T}         \right).
\label{eq:ratio}
\end{equation}
Our conclusion is that the maximum convective flux that we and \cite{Ivanova2018} use are approximately the same.

Despite this similarity \cite{Ivanova2018} concludes that the convection cannot carry the recombination energy out based on her equation (11) and figure 5. However, according to figure 5 of \cite{Ivanova2018} the convective flux is much lower than the recombination flux only where the ionization fraction of hydrogen is $\chi \la 30 \%$. But as we discussed in section \ref{subsec:diffusion}, from that zone the photon diffusion time is much shorter than the dynamical time, namely, less than half a year. Even from the deeper zone where $\chi = 40 \% $ the photon diffusion time is less than the envelope expansion time of $1.7 \yr$.
\cite{Ivanova2018} takes the region where the ionization fraction is $\chi=20 \%$ and claims that the convection flux is too low to transport the recombination energy.  However, from that region the photon diffusion time is very short and there is no problem to carry the energy out just by radiation. If convection is included, the energy transport time will be even shorter. We discuss this further below.

\subsection{On the energy transport in the outer envelope }
\label{subsec:outer}

In her equation (11) \cite{Ivanova2018} compares the maximum convective energy flux with the flux that results from the hydrogen recombination energy. This expression includes two parameters: the ratio of the width of the hydrogen ionization zone to the stellar radius, $\alpha_{\rm H}$, and the ratio of recombination time to dynamical time, $\alpha_{\rm rec}$. For our evolution the average values of these parameters are $\alpha_{\rm H} \simeq 0.2$ and $\alpha_{\rm rec} \simeq 1.7 \yr / 0.8 \yr =2$. When we take the factor $\alpha_{\rm rec}/\alpha_{\rm H} \simeq 10$ and multiply by the other terms of equation (11) of \cite{Ivanova2018} whose values are given in figure 5 of \cite{Ivanova2018}, we find that when the hydrogen ionization fraction is
$\chi \ga 20 \%$ convection alone can carry the recombination energy out.
If we add the radiative flux, which is not much lower than the convective flux in these outer regions (see below), we see no problem for convection and radiation (photon diffusion) to carry the recombination energy out.

We can see it also for the simulations carried out here. We substitute the typical values in equation (\ref{eq:lmaxconv1}) at $t=0.8 \yr$, namely, in the middle of the plunge-in phase, and at the radius where  $\chi \simeq 20 \%$. These values are $r \simeq  360 R_\odot$, $\rho \simeq 3 \times 10^{-9} \g \cm^{-3}$, $\beta \simeq 10$, and $c_s^2 = 9 \km \s^{-1}$. We find that the maximum convective luminosity is $L^{20}_{\rm max,conv,rec} \simeq 4.5 \times 10^4 L_\odot$. Replacing the maximum allowed convective velocity, namely the sound speed, by the convective velocity itself $v_{\rm conv} = 4 \km \s^{-1}$, we find the maximum luminosity there to be $\simeq 2 \times 10^4 L_\odot$. This is still larger than the recombination flux in our simulation $L_{\rm rec, H} \simeq 8800 L_\odot$ (section \ref{subsec:luminosity}).
So convection alone can carry the recombination energy out from regions where $\chi \ga 20 \%$.

Indeed, \cite{Ivanova2018} noticed the problems with convection in the zone where the hydrogen ionization fraction is $\chi \la 20 \%$. But in this outer zone the photon diffusion time is shorter than the convective transport time (Figs. 6-8 in \citealt{Gricheneretal2018}), and most of the extra recombination energy can be transported out by radiation alone. Again, we do not argue that all the recombination energy is radiated away during the plunge-in phase, but rather that most of it is. We estimate that only about 10 per cent of the hydrogen recombination energy, i.e., $f_\gamma\approx 0.1$ in equation (\ref{eq:Fgamma}), and only of the outer envelope, is channeled to envelope expansion and removal during the plunge-in phase.

Let us now justify the usage of the photon diffusion time in the discussion at the end of section \ref{subsec:cells}.
We note that in the above discussion we take the energy transport time in the outer envelope regions where the hydrogen ionization fraction is $20 \% \la \chi \la 40 \%$ to be the photon diffusion time, even if in these regions the local convective flux is larger than the local photon diffusion flux (radiative flux). The reason is that we compare the total time for energy to be transfered out, and not the local flux.
From the outer regions of the envelope, if energy would have been transfered out only by photon diffusion the energy transport time would have been only few times longer than if it would have been carried all the way out by convection only. Namely, photon diffusion is non-negligible. In reality, energy is carried for a short distance mainly by convection, and then mainly be radiation, and the real energy transport time is shorter than both the convective-only and radiative-only energy transport time scales, as expressed in equation (\ref{eq:t(tran)}). Therefore, by only taking the photon diffusion time we actually use a longer energy transport time scale than what a full energy transport calculation would give, and hence \textit{we underestimate the fraction of the recombination energy that is radiated away.}
For that, the argument of \cite{Ivanova2018} that we cannot use the photon diffusion time in these outer parts of the envelope does not hold for the goals of our study. We simply take the pessimistic approach and show that even if we take the photon diffusion alone we still conclude that most of the hydrogen recombination energy is radiated away.

\section{SUMMARY}
\label{sec:summary}

We continued the analysis of our numerical calculations presented in \cite{Gricheneretal2018} regarding the role of hydrogen recombination in the ejection of the envelope in CEE.
Following the criticism by \cite{Ivanova2018} of our \citep{Gricheneretal2018} claim that during the CEE most of the recombination energy of hydrogen is radiated away, we further compared our claims with those of \cite{Ivanova2018}.
The argument circles around the two decades old dispute of whether the recombination energy contributes to common envelope removal (e.g., \citealt{Hanetal1994}) or not (e.g., \citealt{Harpaz1998}).
 The key to understand our claim is to realize that in the outer parts of the envelope where convection becomes inefficient in transporting energy out, photon diffusion becomes very efficient (section \ref{subsec:diffusion} and point 2 below).
 In our previous paper \citep{Gricheneretal2018} we injected energy in the inner part of the envelope of an AGB stellar model to mimic the initial spiraling-in phase of the CEE (termed plunge-in or dynamical phase). By that we inflated the envelope by more than a factor of two in radius within 1.7 years, about the dynamical time of the giant (Fig. \ref{fig:evolution}). In the present study we further analyzed the properties of the inflated envelope.

We presented five points of disagreement in the five subsections \ref{subsec:phase}-\ref{subsec:outer}. We can summarize these as follows.
\begin{enumerate}
\item We simulate the plunge-in phase and not the self-regulated phase of the CEE (that lasts longer), unlike what \cite{Ivanova2018} attributed to us.
\item We showed that from the region where hydrogen is ionized to a degree of $\chi \la 30 \%$, the photon diffusion time out from the envelope is less than a third of the plunge-in time. This implies that radiation by itself, before we add convection, can carry about $ \ga 25 \%$ of the recombination energy of hydrogen.  When we included convection, we found that less than about $20 \%$ of the hydrogen recombination energy can be used to eject the envelope, with a more typical value of $10 \%$.    Our finding contradicts the claim of \cite{Ivanova2018} that the amount of recombination energy that can be transfered away by radiation is negligible.
\item We used the \texttt{MESA} stellar evolution code to follow the inflation of the envelope. \texttt{MESA} uses the mixing length theory and, as we showed in the upper panel of Figs. \ref{fig:structure08} and \ref{fig:structure17}, convection carries most of the recombination energy out. This refutes the claim of \cite{Ivanova2018} who used the mixing length theory to argue that convection can transport out only a negligible amount of the recombination energy.
\item In section \ref{subsec:cells} we elaborated on the differences between the analysis of \cite{Ivanova2018} and our analysis regarding the convective flux. We found that both \cite{Ivanova2018} and we derive approximately the same maximum convective flux (equation \ref{eq:ratio}), and that convection can carry most of the recombination energy in inner regions where the ionization degree is $\chi \ga 20 \%$, out. \cite{Ivanova2018} claims that in regions where the hydrogen ionization fraction is $\chi \la 20 \%$ convection cannot carry energy out, and hence recombination energy will be used to eject the envelope. But, as we already discussed in point (2) above (section \ref{subsec:diffusion}) in those regions photon diffusion carry most of the recombination energy out.
\item Finally, section \ref{subsec:outer} we showed that for our model  convection can carry the recombination energy when the ionization fraction of hydrogen is $\chi \ga 20 \%$. In the zones farther out, where $\chi \la 20 \%$, the photon diffusion time is shorter than the convection time, and therefore most of the recombination energy is carried out by photon diffusion. We also justified our usage of the photon diffusion time in the regions where the ionization fraction of hydrogen is $20 \% \la \chi \la 40 \%$, despite the fact that  the convection flux is larger than the radiative flux there. This is justified for the goals of the present study, contrary to the claim of \cite{Ivanova2018}. By including also convection, we would have found that more recombination energy leaks out and radiates away. Namely, by using the photon diffusion time alone we underestimate the fraction of the recombination energy that is radiated away.
\end{enumerate}

In this study we considered only the hydrogen recombination energy. The two helium recombination zones, that of He$^{++}$ and of He$^{+}$, are much deeper in the envelope. The energy transport times from these zones to the surface are longer than the duration of the plunge-in phase. However, the location of these zones makes the recombination energy less efficient in boosting mass loss. The CEE starts with the envelope inflation as the secondary star spirals-in deep into the giant envelope. Hydrogen recombination takes place in the outer zone with little mass above the recombination zone. If all this energy could have been used (and we showed it cannot), then it would have supplied the extra energy that is required to eject the very outer zone of the now inflated envelope. The recombination energy of helium is released much deeper in the envelope, and its energy should remove a much larger mass. Its energy is actually added to the orbital energy that is also released deep inside the envelope.
So, our expectation is that the effect of the recombination energy of helium is small, and its effect is comparable to the secondary star in-spiraling a little deeper into the envelope.
Namely, the recombination energy of helium will increase somewhat the envelope inflation, but will not cause a massive outflow.
Eventually, as the spiral-in process slows down, there will be enough time to transport the recombination energy of helium out.

We considered here only low mass AGB stars. When we turn to massive stars and/or smaller stars, like sub-giant branch stars, the envelope binding energy is larger, and recombination energy plays an even smaller role.

We summarize by reiterating our previous results that during the CEE most of the hydrogen recombination energy is radiated away (about 90 per cent). The recombination energy does not contribute much to the energy that is used to remove and accelerate the envelope.
Our results show that the inclusion of recombination energy in the common envelope simulation should only be considered once radiation transport is included in the codes, or else the dynamics of the CEE simulation would be erroneously altered.
It is our view that in cases where extra energy  sources to the orbital energy are required it is more likely that the companion star contributes the energy by accreting envelope mass and launching jets.

\section*{Acknowledgments}
We thank an anonymous referee for many detailed and thoughtful comments throughout the reviewing process that helped us clarify some unclear explanations and improved the manuscript. We thank Natasha Ivanova for detailed comments.
We acknowledge support from the Israel Science Foundation and a grant from the Asher Space Research Institute at the Technion. A.G. was supported by The Rothschild Scholars Program- Technion Program for Excellence.





\label{lastpage}
\end{document}